\documentclass{article}
\usepackage{ijcai22}
\usepackage{natbib}
\usepackage{amsfonts,amssymb}

\usepackage{tikz}
\usepackage{pgfplots}
\usepackage{times}

\usepackage{soul}

\usepackage[hidelinks]{hyperref}
\usepackage[utf8]{inputenc}
\usepackage[small]{caption}
\usepackage{graphicx}
\usepackage{amsmath}
\usepackage{booktabs}
\usepackage{multirow}
\usepackage{multicol}
\usepackage{listings}
\lstdefinestyle{Python}{
	language        =   Python, 
	basicstyle      =   \zihao{-5}\ttfamily,
	numberstyle     =   \zihao{-5}\ttfamily,
	keywordstyle    =   \color{blue},
	keywordstyle    =   [2] \color{teal},
	stringstyle     =   \color{magenta},
	commentstyle    =   \color{red}\ttfamily,
	breaklines      =   true,   
	columns         =   fixed,  
	basewidth       =   0.5em,
}

\title{Non-autoregressive Model for Full-line Code Completion}

\newcommand\correspondingauthor{\thanks{Corresponding authors.}}

\author{
Fang Liu$^{1,2}$\and
Zhiyi Fu$^{1,2}$\and
Ge Li$^{1,2}$\correspondingauthor \and
Zhi Jin$^{1,2*}$ \and
Hui Liu$^3$\and
Yiyang Hao$^4$\\
\affiliations
$^1$Key Laboratory of High Confidence Software Technologies (Peking University), Ministry of Education\\
$^2$ School of Computer Science, Peking University, Beijing, China \\
$^3$Beijing Institute of Technology\\
$^4$Silicon Heart Tech Co., Ltd\\
\emails
\{liufang816, ypfzy, lige, zhijin\}@pku.edu.com,
liuhui08@bit.edu.cn,
haoyiyang@nnthink.com
}

\begin{document}

\maketitle

\begin{abstract}
Code completion tools are frequently used by software developers to accelerate software development by suggesting the following code elements. Completing a sequence of code tokens (e.g., a full line of code) has been proved more efficient than predicting a single token at a time.  To complete the code sequence, researchers are employing AutoRegressive (AR) decoders to generate tokens in a left-to-right, token-by-token fashion. Consequently, the prediction of the next token depends on all previously generated tokens, which leads to high latency in inference. To improve the efficiency and accuracy of full-line code completion, in this paper, we propose a Non-AutoRegressive (NAR) model for code completion boosted by a syntax-aware sampling strategy. Our experimental results on two widely used datasets suggest that our model outperforms both AR and NAR baselines on full-line code completion, and it is faster than the AR model with up to $9\times$ speed-up.
\end{abstract}

\section{Introduction}
Code completion is one of the most useful features in the Integrated Development Environments (IDEs), improving the software development efficiency by suggesting future code snippets.  
In recent years, as the development of machine learning and deep learning technologies and easy-to-acquire open-source codebases, researchers have started to tackle code completion by learning from large-scale code corpora. Various Language Models (LM) have been employed for code completion, including N-gram \citep{hindle2016naturalness}, RNN \citep{li2017code}, and Transformer based models \citep{liu2020self}.
Several popular industry code completion tools, e.g., \cite{tabnine}, \cite{aiXcoder}, and \cite{Copilot}, also rely on LMs to provide suggestions for whole lines or entire functions in a token-by-token manner.
Recently, \cite{wang2020towards} further explored full-line code completion. Given a partially completed code snippet, they built a Transformer-based LM to predict the next line of code.
Although such code completion algorithms/tools can suggest longer code snippets, they intrinsically are suffering from inefficiency because of the autoregressive decoding process, where each token is generated conditioned on all the previously generated tokens. 
Consequently, the generation is not parallelizable and thus particularly slow. 
However, \textbf{the efficiency of the generation process is of great importance for code completion} because it is expected to respond instantly on the developer’s own devices.

Tokens within a code line have the potential to be predicted concurrently, although existing code completion algorithms predict them one by one (from left to right) by exploiting all tokens previously predicted. 
For example, multiple arguments of the same method are independent of each other, and thus they could be predicted independently and concurrently. In Python code, you can even change the order (position) of the arguments if a function with the keyword `arguments' is called. For example, \verb|Func(arg1=a, arg2=b)| is equivalent to \verb|Func(arg2=b, arg1=a)|.
We also argue that even if there exists strong dependency among tokens within a code line, the dependency is not necessarily left-to-right. 
A typical example is ``one line IF-ELSE statement'' (Python), formed as: \verb|value_1 if condition else value_2|.  
The $value\_1$ is dependent on the following condition (on its right-hand side), instead of the verse.  We also perform in-depth analysis in section \ref{motivation} to verify that the dependency among code tokens is weaker than that among tokens in natural languages where parallelized generation has been employed successfully.  Our analysis in section \ref{motivation} also suggests that predicting code tokens from right-to-left or the verse (left-to-right) results in comparable performance. Based on those observations, we conclude that generating tokens in parallel for code completion is possible and reasonable.

Non-autoregressive (NAR) model, proposed by \cite{gu18nat}, has been successfully applied to Natural Machine Translation (NMT) to generate tokens non-autoregressively through a parallel decoding process:
\begin{equation}
    \mathcal{L}_{NAR} = \sum_{t=1}^{T} \log p(y_t|X;\theta)
\end{equation}
It allows an order of magnitude lower latency during inference. There are two kinds of NAR models: Fully NAR models \citep{gu18nat,Ma19FlowSeq} and iterative NAR models \citep{Ghazvininejad2019mask-predict,kasai2020non}. These NAR models can speed up the inference process of NMT significantly compared with autoregressive models. However, NAR models are hard to train without specially designed training strategies, as the left-to-right inductive bias among target tokens is ignored during decoding.

Considering the feasibility of parallel generation of code and inspired by the success of Non-AutoRegressive generation in NLP, in this paper, we propose a \textbf{S}yntax-\textbf{A}ware \textbf{N}on-\textbf{A}uto\textbf{R}egressive model (\textbf{SANAR}) for code completion. Specifically, we propose an adaptive and syntax-aware sampling strategy to boost the training process of SANAR, which dynamically glances some code tokens from the target sequence according to their difficulties and token types. The better the model is trained, the fewer code snippets would be glanced. Once the model is well-trained, it can generate the whole line of tokens in a single pass. Our experimental results show that our sampling strategy results in obvious improvement in performance on both Python and Java code completion. The proposed approach is significantly faster than the widely used Auto-Regressive models, achieving $5\sim 6\times$ speed-up on average, and $9\times$ for long targets.

To summarize, the major contributions of our work are as follows:

\begin{itemize}
    \item An empirical study whose results suggest that it is potentially practical to predict code tokens in parallel. 
    \item A novel approach for full-line code completion. To the best of our knowledge, it is the first non-autoregressive approach to code completion. We boost the approach by an adaptive and syntax-aware sampling strategy, that is specially designed for source code. 
    \item A large-scale evaluation of the proposed approach whose results suggest that our approach outperforms both AR and NAR baselines on full-line code completion, and significantly reduces the inference time compared with the AR model. 
\end{itemize}

\section{Dependency among Generated Code Tokens}\label{motivation}
We argue that the dependency among the generated tokens in full-line completion is not as strong as in NMT (where NAR has been successfully applied), and the left-to-right generation order is not always optimal for code completion. To verify this assumption, we conduct experiments to analyze the dependency among target tokens in the full-line code completion task in this section.

\begin{table}[t]
\centering \small
\setlength{\abovecaptionskip}{0cm} 
\caption{The percentage of the code sequence that can be correctly handled by different models. EM stands for exact match, and ES stands for edit similarity.}
\begin{tabular}{l|l|l|c}  
\toprule
 ~ & ~ & Metrics & Percentage  \\
\midrule
\multirow{6}{*}{PY} & \multirow{2}{*}{only Transformer-L2R} & EM  & 2.65\%   \\
& ~ & ES\textgreater 0.5 & 6.57\%  \\
& \multirow{2}{*}{only Transformer-R2L} & EM & 2.47\% \\
& ~ &  ES\textgreater 0.5 & 7.21\% \\
& \multirow{2}{*}{Both} & EM & 13.59\% \\
& ~ & ES\textgreater 0.5 & 60.39\% \\
\midrule
\multirow{6}{*}{JAVA} & \multirow{2}{*}{only Transformer-L2R} & EM  & 3.28\%   \\
& ~ & ES\textgreater 0.5  & 6.71\%  \\
& \multirow{2}{*}{only Transformer-R2L} & EM & 3.86\% \\
& ~ &  ES\textgreater 0.5 & 8.22\%  \\
& \multirow{2}{*}{Both} & EM  & 26.15\% \\
& ~ & ES\textgreater 0.5  & 60.43\% \\
\bottomrule
\end{tabular}
\vspace{-0.2cm}
\label{tab:order_percentage}
\end{table}

\subsection{Reversing the Order of Code Generation}
To verify our conjecture of left-to-right is not always the optimal order for code completion, we conduct an experiment to analyze the impact of different generation orders. We employ standard auto-regressive Transformer architecture to perform full-line code completion using two reverse orders: left-to-right (Transformer-L2R) and right-to-left (Transformer-R2L).
Table \ref{tab:order_percentage} presents the results. For most of the cases, the performances of the L2R model and R2L model are comparable. For Python language, 2.65\% and 2.47\% of programs can only be correctly generated by L2R and R2L, respectively; 6.57\% and 7.21\% of the programs can only be approximately generated (edit similarity \textgreater 50\%) by L2R and R2L, respectively. For Java programs, we can also get similar results. Therefore, we conclude that left-to-right is not always the optimal order for code completion. Thus generating code in parallel (non-autoregressively) could be a better choice.

\begin{figure}[t]
\setlength{\abovecaptionskip}{0cm}
\footnotesize
    \begin{flushleft}
    \begin{tikzpicture}
        \begin{axis}[width=0.85\columnwidth,
                     height=0.5\columnwidth,
                     xlabel=P,
                     ylabel=R(p),
                     ymin=0.55,ymax=0.7,
                     legend pos= north east,
                     legend style={font=\tiny},
                     ymajorgrids=true,
                     grid style=dashed]
\addplot[draw=blue,mark=diamond*] coordinates {(0.15,0.59)(0.35,0.5832)(0.5,0.58)};
\addlegendentry{Code Completion}
\addplot[draw=red,mark=square*,color=red] coordinates {(0.15,0.63)(0.35,0.61)(0.5,0.60)};
\addlegendentry{NMT}
\end{axis}
\end{tikzpicture}
\end{flushleft}
\caption{The attention density ratio R(P) under different masking probability P in code completion and NMT.}
\label{Fig:dependency_res}
\vspace{-0.3cm}
\end{figure}
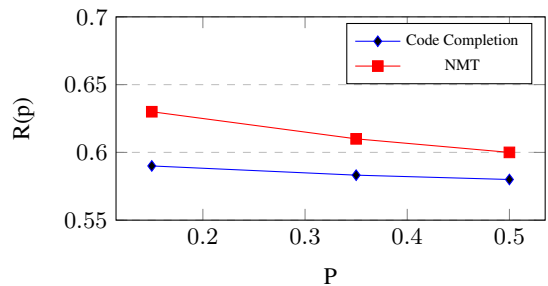

\begin{figure*}[t]
    \centering
    \setlength{\abovecaptionskip}{0.1cm} 
    \includegraphics[width=0.7\linewidth]{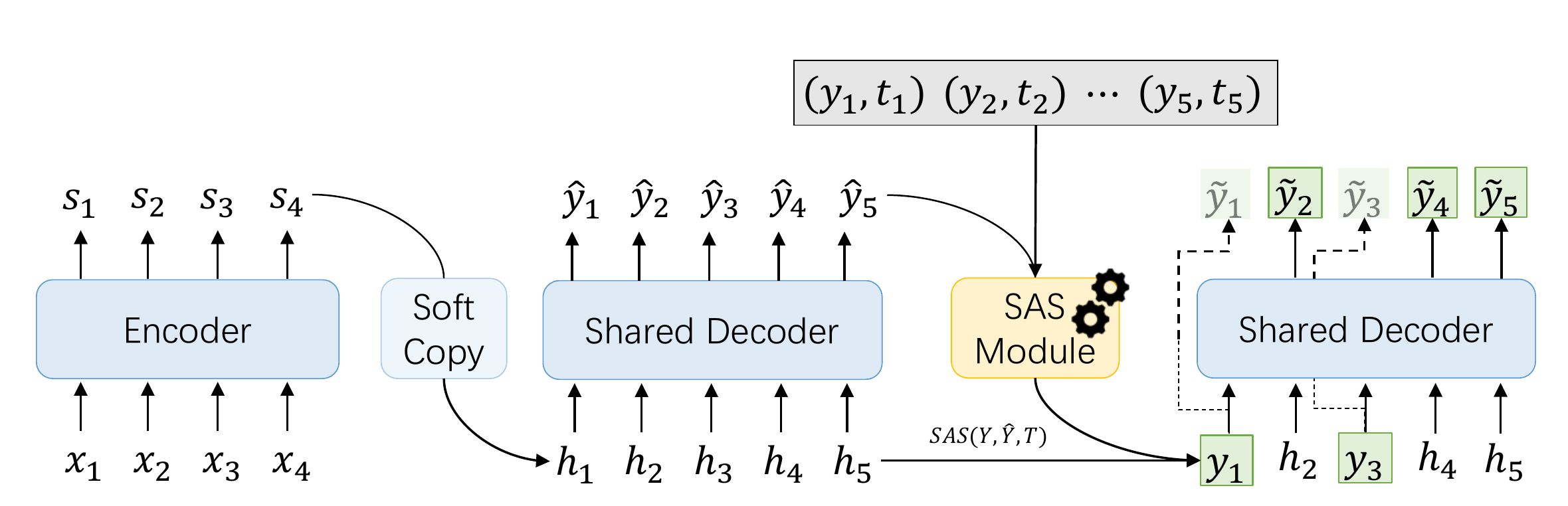}
    \caption{The training procedure with syntax-aware sampling strategy in SAS.}
    \vspace{-0.3cm}
    \label{fig:model}
\end{figure*}

\subsection{Quantitative Dependency among Code Tokens}
In this section, we measure the quantitative dependency among the generated tokens by attention density and compare the dependency among code tokens against that among natural language tokens (in NMT tasks). NMT task is selected for comparison because NAR models have been successfully applied to it. 

To characterize and quantify the target-token dependency, inspired by \cite{Ren20Study}, we build Dependency-Analysis-Model (DAM) to measure the dependency among target tokens using the ratio of the attention weights on target context over that on full (both source and target) context when predicting a target token, where \textbf{a bigger ratio indicates a larger dependency among target tokens}. Specifically, we use masked language modeling task \citep{Devlin19bert} to train DAM. 
We employ mix-attention \citep{he2018layer} to calculate the attention weights,
where source tokens can only attend to source tokens with self-attention and target tokens can attend to all source and target tokens with mix-attention. During training, we randomly mask the tokens on the target side, and the model is trained to predict the original value of these tokens. 
After the model has been trained, we can measure the target token dependency based on the ratio of attention weights $\alpha_i$ on target context over that on full context when predicting a specific target token $y_i$, which is defined as: 

\begin{equation}
    \alpha_i = \frac{\frac{1}{N} \sum_{j=1}^{N} A_{i,j}}{\frac{1}{N} \sum_{j=1}^{N} A_{i,j} + \frac{1}{M} \sum_{j=N+1}^{N+M} A_{i,j}}
\end{equation}

\noindent where $A_{i,j}$ denotes the attention weights from token $i$ to token $j$ in mix-attention, and $j \in [1,N]$ represents the target token, and $j \in [N+1,N+M]$ represents the source token. $M$ and $N$ is the length of source and target input, respectively. $\sum_{j=1}^{N+M} A_{i,j} = 1$. $\alpha_i$ represents the ratio of attention density on target context when predicting target token $i$. For a given masking probability $P$, the final attention density ratio $R(P)$ is calculated by averaging $\alpha_i$ over all test data. As seen in Figure \ref{Fig:dependency_res}, the attention density ratio for NMT is bigger than code generation for all masking probability $P$, which demonstrates the dependency among the target tokens in code completion is less than NMT.

We conclude based on the preceding analysis that it is potentially practical to predict code tokens in parallel.

\section{SANAR}

\subsection{Overview}

In this section, we present our model SANAR in detail. We build SANAR to perform full-line code completion in parallel, which uses conditional independent factorization for the target tokens. Specifically, given the contextual code sequence $X=\{x_1,x_2,...,x_m\}$, the non-autoregressive full-line code completion task is to predict a sequence of tokens  $Y=\{y_1,y_2,...,y_n\}$ that form a complete statement in a non-autoregressive way:
\begin{equation}
    P(y_1,...,y_n|x_1,x_2,...,x_m) = \prod_{t=1}^n P(y_t|x_1,x_2,...,x_m)
\end{equation}
Figure \ref{fig:model} shows the architecture and the trainig procedure of our model. SANAR adopts the encoder-decoder transformer architecture: a context encoder that does self-attention, and a generated-code decoder that has one set of attention heads over the encoder’s output and another set (self-attention) for the generated code. During the training procedure, we propose an adaptive syntax-aware sampling strategy, which enable SANAR to generate target code sequences in a single-pass during inference procedure by gradual training. 

\subsection{Training}
During training, SANAR adopts an adaptive Syntax-Aware Sampling (SAS) strategy to dynamically glance code snippets from the target sequence depending on its difficulty and the tokens' syntax types, aiming to incorporate the explicit syntactic information of the program. To achieve the sampling strategy, SANAR performs decoding twice during training. As shown in Figure \ref{fig:model}, the first decoding is performed in a fully-NAR way, where the input to the decoder $H =\{h_1, h_2,...,h_n\}$ are copied from the encoder output using soft-copy \citep{Wei19Imitation}, which maps the encoder embeddings $S = \{s_1, s_2,...,s_m\}$ into target length $H = \{h_1, h_2,...,h_n\}$ depending on the distance relationship between the source position $i$ and the target position $j$. The initial predicted tokens $\hat{Y}$ are predicted as:

\begin{equation}
    \hat{Y} = f_{dec}(\text{soft-copy}(f_{enc}(X;\theta'));\theta)
\end{equation}

\noindent The prediction accuracy indicates the difficulty of fitting current target. In the second decoding, SANAR samples words of the targets as the extra decoding input by SAS sampling according to the first decoding results and the syntax information of the target tokens, and learn to predict the rest words that are not selected. It is important to note that only the second decoding will update the model parameters. 

The training objective is to maximize the following:
\begin{equation}
    \mathcal{L}_{SAS} = \sum_{y_t \in Y \backslash \mathbb{SAS}(Y,\hat{Y}, T)} \log P(y_t|X, \mathbb{SAS}(Y,\hat{Y}, T);\theta)
\end{equation}

\noindent where $\hat{Y}$ is the initially predicted tokens in a fully-NAR decoding way. $T$ is the syntax type sequence of the target sequence, 
and $\mathbb{SAS}(Y,\hat{Y}, T)$ is a subset of tokens selected via the adaptive SAS strategy. 

\subsection{Syntax-Aware Sampling Strategy}
Syntax-Aware Sampling (SAS) strategy is an important component of SANAR, which adaptively selects the positions of tokens from the target sequence depending on its difficulty and the syntax types of the tokens. The selected tokens can provide ``correct” information from the ground-truth target, thus can reduce the burden of decoder to predict the rest non-selected tokens in the training phase. SAS samples more tokens for SANAR to glance at the beginning, and then reduces the sampling number gradually, which helps SANAR to learn, eventually, how to predict the whole line code snippet without seeing any ground truth tokens in one pass.

Formally, given the context code sequence $X$, its predicted token sentence $\hat{Y}$, the ground truth $Y$, and the token's syntax type sequence of the target $T$, the goal of SAS strategy $\mathbb{SAS}(Y,\hat{Y}, T)$ is to obtain a subset of tokens sampled from $Y$. Specifically, there are two steps:

\noindent 1) Deciding the sampling numbers $N$ depending on the difficulty of correctly generating the target token sequence:
\begin{equation}
    N = \lambda \cdot dis (Y, \hat{Y})
\end{equation}
$N$ is computed by comparing the difference between $\hat{Y}$ and $Y$, and we adopt the Hamming distance as the metric following \cite{Qian21glat}, $dis(Y, \hat{Y}) = \sum_{t=1}^T(y_t \neq \hat{y_t})$. $\lambda$ is the sampling ratio, which is a hyper-parameter to flexibly control the number of sampled tokens. More tokens will be selected and fed as input of the second-pass decoding if the network’s initial prediction is less accurate. Thus, the sampling number can be decided adaptively considering the current trained model’s prediction capability and the training sample's complexity. 

\noindent 2) Sampling $N$ tokens from the target sequence. The most direct method is to randomly select $N$ tokens from $Y$ like in \cite{Qian21glat}. However, we argue that considering the syntactic information of the code explicitly during token selection will be beneficial for understanding the programs, and thus can help train the decoder to predict the rest tokens more precisely. In programs, the keyword, identifiers, and operators contain more symbolic and syntactic information than other tokens (for example, literals and separators in Java, like `;', `\{', `\}', etc.). Glancing these tokens will help a lot. 

To capture the syntactic information of the code during the sampling procedure, we present a Hybrid-Syntax-Guided (HSG) sampling strategy. Specifically, $1-p$ of the time, the sampling is performed randomly, where the tokens are randomly selected from the target sequence; and $p$ of the time, tokens are selected depending on their syntax-types\footnote{$p$ is a hyper-parameter to control the probability of syntax-guided sampling, the best setting is $p=30\%$}, we randomly select $K\leq N/2$ keywords, $I \leq N/4$ identifiers, and $O \leq N/4$ operators from $Y$, where $K+I+O \leq N$\footnote{The number of these specific tokens might be not enough}. The reason for introducing randomness (randomly sampling for $1-p$ of the time) in the sampling process is to enable SANAR to explore more interdependency among target tokens. Compared with the totally randomly sampling strategy in \cite{Qian21glat}, our Hybrid-Syntax-Guided sampling strategy can increase the probability of glancing the keyword, operators, and identifiers, which can help the SANAR capture the program's syntactic and semantic information better.

Finally, we can obtain a subset of words sampled from $Y$:
\begin{equation}
\begin{split}
    \mathbb{SAS}(Y,\hat{Y}, T) &= HSG(Y, N, p) \\
    N &= \lambda \cdot dis (Y, \hat{Y})
\end{split}
\end{equation}

\subsection{Inference}
As SANAR is well-tuned, it will adaptively reduce the percentage of sampling, making sure that the trained model could learn to generate the whole line of code in the single pass. Thus, the inference of SANAR is fully parallel with only a single pass. 

In traditional left-to-right code completion, where the target sequence is predicted token by token, the length of the sequence can be determined by predicting a special \textsc{EOS} (end of sentence) token. However, for NAR-based generation where the entire sequence is predicted in parallel, the length of the target sequence must be determined in advance. To achieve this, we follow \citet{Ghazvininejad2019mask-predict} to add an additional \texttt{[LENGTH]} token to the source input, and the encoder output for the \texttt{[LENGTH]} token is used to predict the length, formed as a \textit{max-target-length} classification task. We use 1,000 as the max length. The loss is added to the cross-entropy loss from the target sequence.

\section{Experiments}

\subsection{Experimental Settings}
\noindent\textbf{Datasets.}
We conduct experiments on two program benchmarks crawled from Github repositories: Py150 \citep{Raychev16Probabilistic} contains 150,000 Python files, split into a training set of 100,000 files and test set of 50,000 files, GitHub-Java \citep{Allamanis13Mining} contains 14,317 projects, we follow \cite{hellendoorn2017deep} and \cite{karampatsis2020big} to split validation and test set, and randomly sample 1,000 projects from the rest projects as the training set. 

We use the Python official library tokenizer\footnote{https://docs.python.org/3/library/tokenize.html} and Javalang\footnote{https://github.com/c2nes/javalang} to split Python and Java programs into lines of tokens, and extract their syntax types. Then we employ a sliding context window of 10-lines to create the data pairs, that is, the context code sequence contains 10-lines of code tokens, and the next line is considered as the target code sequence. The detailed information of the datasets is shown in Table \ref{tab:dataset}.

\begin{table}[]
    \centering \small
    \setlength{\abovecaptionskip}{0.1cm} 
    \caption{Dataset statistics.}
    \begin{tabular}{l|c|c}
    \toprule
         & Python & Java  \\
    \midrule
    \# of training pairs  & 7,531,208 & 12,993,112 \\
    \# of testing pairs  & 3,693,213 & 671,236\\
    Avg. tokens in cxt  & 90.8 & 71.2 \\
    Avg. tokens in tgt  & 9.4 & 6.9 \\
    \bottomrule
    \end{tabular}
    \vspace{-0.2cm}
    \label{tab:dataset}
\end{table}

\noindent\textbf{Metrics and Baselines.}
We use the following metrics to evaluate the performance of our approach:
\begin{itemize}
    \item Exact Match Accuracy (EM): We compare the exact matching accuracy between the generated code sequence and the ground truth.
    \item BLEU: We use the BLEU score to measure the n-gram similarity between the target code sequence and the generated code sequence.
    \item Edit similarity (ES): We use the character-level edit similarity of the predicted output $\hat{Y}$ and the target output $Y$, which is computed as:
    \begin{equation}
        ES = 1 - \frac{Lev(\hat{Y}, Y)}{|\hat{Y}| +|Y|}
    \end{equation}
    where \textit{Lev} is Levenshtein distance.
    \item Latency: The inference latency is computed as the time to decode a single target code sentence without mini batching, averaged over the whole test set. The decoding is implemented in PyTorch on a single NVIDIA Tesla V100.
\end{itemize}

\noindent We compare our model with the following baselines, including both strong AR Transformer and several representative NAR models:
\begin{itemize}
    \item Transformer \citep{vaswani2017attention}: Base Transformer seq2seq architecture, where the decoder is \textbf{AR}. We also tried the TransformerLM decoder architecture in \cite{wang2020towards} using the comparable number of parameters. The results show that seq2seq architecture can outperform the LM decoder in performance, and with a comparable decoding speed. Thus we use Transformer (seq2seq) model as the strong AR baseline, and also use seq2seq framework as our base architecture instead of LM decoder.
    \item CMLM \citep{Ghazvininejad2019mask-predict}: A strong \textbf{iterative-NAR} model, which adopts multi-pass decoding strategy to refine the target tokens.
    \item GLAT \citep{Qian21glat}: A strong \textbf{fully-NAR} model, which can generate high-quality target only with single-pass parallel decoding.
\end{itemize}

\noindent\textbf{Setup.}
For all the seq2seq baselines and our model, we use a 6-layers of encoder and decoder with a model size of 512\footnote{For TransformerLM \citep{wang2020towards}, to keep the comparable number of the parameters, we employ a 12-layers decoder with a model size of 512.}.
We set the vocabulary size to 50,000 for both Python and Java datasets.  We train the model with batches of 16k tokens using Nvidia V100 GPUs. We set the dropout rate to 0.1 and use Adam optimizer with $\beta = (0.9, 0.999)$. The learning rate warms up to 5e-5 in 4k steps and gradually decays according to the inverse square root of the step number \citep{vaswani2017attention}. Following \cite{Qian21glat}, we set $\lambda$ to 0.3. For the hyper-parameter $p$ in Hybrid-Syntax-Guided sampling strategy, we use 30\% as the final value, which can achieve the best results. The specific syntax types used for guiding the sampling include keyword, operator, and identifier.

\begin{table}[t]
    \centering
    \small
    \setlength{\abovecaptionskip}{0.1cm} 
    \caption{Results on Python benchmark.}
    \begin{tabular}{l|c|c|c|c|c}
    \toprule
    Model & BLEU & EM & ES & Latency & Speedup \\
    \midrule
    Transformer & 21.93 & 16.24 & 62.73  & 121ms & $1.0\times$ \\
    \midrule
    CMLM & 23.98 & 13.44 & 63.82 & 39ms & $3.1\times$\\
    GLAT & 26.78 & 16.95 & 66.36 & \textbf{20ms} & $6.1\times$\\
    \midrule
    SANAR & \textbf{29.07} & \textbf{18.35} & \textbf{66.90} & \textbf{20ms} & $6.1\times$ \\
    \bottomrule
    \end{tabular}
    \vspace{-0.2cm}
    \label{tab:py_res}
\end{table}

\begin{table}[t]
    \centering
    \small
    \setlength{\abovecaptionskip}{0.1cm} 
    \caption{Results on Java benchmark.}
    \begin{tabular}{l|c|c|c|c|c}
    \toprule
    Model & BLEU & EM & ES & Latency & Speedup \\
    \midrule
    Transformer & 25.73 & 30.33 & 63.95 & 101ms & $1.0\times$ \\
    \midrule
    CMLM & 28.97 & 28.36 & 63.66 & 31ms & $3.3\times$\\
    GLAT & 30.56 & 28.63 & 65.80 & \textbf{20ms} & $5.1\times$ \\
    \midrule
    SANAR & \textbf{32.41 }& \textbf{30.57} & \textbf{66.98} & \textbf{19ms} & $5.3\times$ \\
    \bottomrule
    \end{tabular}
    \vspace{-0.2cm}
    \label{tab:java_res}
\end{table}

\subsection{Main Results}
The main results on Python and Java benchmarks are presented in Table \ref{tab:py_res} and \ref{tab:java_res}. SANAR significantly improves the full-line code completion quality and outperforms all the baselines on all evaluation metrics. 
Especially, SANAR outperforms strong AR baseline (Transformer) by a large margin in BLEU. We can observe that the improvements on BLEU and Edit Similarity are more significant than the Exact Match accuracy. As a recommendation tool, code completion add-in serves as developers' cooperator, and developers can accept to make a few edits to correct the recommended code. Besides, different code snippets can have identical functionality. For example, these two Python return statements have the same functionality:
\verb|return True if a==0 else False| and \verb|return False if a!=0 else True|. Only evaluating the quality of the generated code by comparing the exact match with the ground truth is not convincing enough. Thus, BLEU and Edit Similarity which calculate the token overlapping and the minimum number of operations required to transform the predicted code sequence into the target sequence can evaluate the generated code more thoroughly. Thus, the significant improvements on BLEU and ES further demonstrate that SANAR can generate the code sequence which meets the developers' expect more.

It is interesting to note that, most of the NAR models can achieve higher BLEU and ES scores than the AR-based Transformer model, and can achieve better or comparable EM accuracy on all the datasets. These results are quite different from NMT task, where the NAR models can achieve worse or comparable results with the AR model. The results further verify our assumptions in section \ref{motivation}, that is, the non-autoregressive manner can fit code completion better than NMT, and can boost the performance on both efficiency and quality for code completion.

\begin{table}[]
    \centering \small
    \setlength{\abovecaptionskip}{0.1cm} 
    \caption{Speed-up on long target sequences.}
    \begin{tabular}{l|cc}
    \toprule
     Model & Latency & Speed up \\
    \midrule
    Transformer & 171ms & $1.0\times$ \\
    CMLM & 34ms & $5.0\times$ \\
    GLAT & 19ms & $9.0\times$ \\
    SANAR & 19ms & $9.0\times$ \\
    \bottomrule
    \end{tabular}
    \vspace{-0.2cm}
    \label{tab:long_sequence}
\end{table}

For the inference latency, SANAR can significantly reduce the inference time within $5\sim6\times$ speedup compared with the AR-based Transformer model. As a fully-NAR model, the inference speed of SANAR is also faster than the iterative-NAR CMLM model. Compared with NMT, the target sequence in full-line code completion is shorter, thus the overall speedup is not as large as in NMT, but is still significant. We also present the latency comparison for completing lines of $\geq10$ tokens, which account for about 30\% of the test set. The results are shown in Table \ref{tab:long_sequence}. As SANAR can generate tokens in parallel, thus the latency will not be affected by the target lengths, and still keeps comparable with previous results. However, for AR models, as the number of generated tokens increases, the latency increases a lot, and SANAR can achieve $9\times$ speedup. Thus, when completing longer token sequences, the speedup of SANAR will be more significant.

\begin{table}[]
    \centering \small
    \setlength{\abovecaptionskip}{0.1cm} 
    \caption{Performance of full-line code completion with different sampling probabilities.}
    \begin{tabular}{l|ccc|ccc}
    \toprule
    \multirow{2}{*}{p} & \multicolumn{3}{c|}{Python} &  \multicolumn{3}{c}{Java} \\
         & BLEU & EM & ES & BLEU & EM & ES  \\
    \midrule
    0 & 26.78 &	16.95 & 65.94 & 30.56 &	28.63 & 65.80 \\ 
    0.15 & 28.20 & 16.96 & 66.36 & 32.16 & 29.94 & 66.81 \\
    0.3 & \textbf{29.07} &\textbf{ 18.35} & \textbf{66.90} & \textbf{32.41}	& \textbf{30.57} & \textbf{66.98} \\
    0.5 & 27.37	& 17.31 & 66.30 & 31.29	& 29.82 & 66.43 \\
    \bottomrule
    \end{tabular}
    \vspace{-0.2cm}
    \label{tab:diff_p}
\end{table}

\begin{table}[]
    \centering \small
    \setlength{\abovecaptionskip}{0.1cm} 
    \caption{Token repetition ratio results.}
    \begin{tabular}{c|cc}
    \toprule
      & Python & Java \\
     \midrule
     Transformer & 3.72\% & 4.36\% \\
     CMLM & 3.66\% & 4.92\% \\
     GLAT & 4.55\% & 4.56\% \\
     SANAR & 5.52\% & 5.40\% \\
     \bottomrule
    \end{tabular}
    \vspace{-0.2cm}
    \label{tab:repetition}
\end{table}

\subsection{Analysis}

\noindent\textbf{Influence of Parameter $p$.} ~ To analyze how the sampling probability $p$ of our Hybrid-Syntax-Guided sampling strategy affects the model's performance, we conduct experiments with different $p$s, where a larger $p$ indicates more tokens are sampled depending on their syntax-types. When $p$ is zero, the sampling will become totally random, which is the same with GLAT \citep{Qian21glat}. The results for different $p$s are listed in Table \ref{tab:diff_p}. When $p$ is increased from 0 to 0.3, the performance becomes better, demonstrating the effectiveness of the syntax-guided sampling. When $p$ is further increased, where the randomness decreases, the performance began to drop, but still outperforms GLAT. This suggests that it is also necessary to introduce randomness during the syntax-aware sampling, which can help SANAR explore more interdependency among target tokens. 

\noindent\textbf{Token Repetition Ratio.} ~ In NMT, one of the known problems in non-autoregressive models is the high token repetition ratio, since the interdependency among target tokens is not considered. We are also interested in whether this problem also troubles code completion. We measure the percentage of repeated tokens on the test set of Python and Java benchmarks. The results are shown in Table \ref{tab:repetition}.  As seen from the results, the token repetition ratio of the NAR-based models is not increased obviously when compared with the AR-based Transformer model, which is quite different from NMT. These results also demonstrate that the left-to-right interdependency among target tokens in the code completion task is much weaker than that of NMT task. Thus, the non-autoregressive decoder can fit code completion better, and achieve better performance.

\section{Related Work}
\subsection{Code Completion}
Since \citet{hindle2016naturalness} have shown that source code is highly repetitive and predictable, language models began to be used for source code modeling and code completion \citep{hindle2016naturalness,tu2014localness,hellendoorn2017deep,li2017code,liu2020self}. Most of the LM-based code completion models perform next one token completion \citep{li2017code,liu2020self}. \cite{wang2020towards} propose to use a Transformer-based LM to perform full-line code completion. In full-line code completion, instead of predicting the next single token, the model predicts a sequence of tokens that form a complete statement given a partially complete code context. The experimental results show that Transformer seq2seq architecture can outperform the LM decoder in \cite{wang2020towards} with comparable number of parameters. Thus, we use Transformer seq2seq architecture as the AR baseline, and also employ seq2seq architecture for SANAR instead of only using a decoder as \cite{wang2020towards}. \cite{guo2021learning} proposed GRAMMFORMER, a transformer-based grammar-guided code generation model that generates code sketches, i.e. sequences of code tokens with holes. As the generation target is different from our model, we did not compare with them. 

\subsection{NAR Models of Text Generation}
\noindent\textbf{Fully NAR} ~ The Fully Non-AutoRegressive model consists of the same encoder as Transformer and a parallel decoder \citep{gu18nat}. During training, it uses the conditional independent factorization for the target sentence by maximizing the following likelihood:
\begin{equation}
    \mathcal{L}_{NAR} = \log P(Y|X;\theta)=\sum_{t-1}^{T} \log p(y_t|X;\theta)
\end{equation}
During inference, the encoder representation is copied as the input for the decoder, therefore all tokens on the target side can be generated in parallel without depending on the previously generated tokens. Recently, \citet{Qian21glat} propose GLAT, which can achieve parallel text generation with only a single decoding pass by gradual training. 

\noindent\textbf{Iterative NAR} ~ The conditional independence assumption in fully NAR does not hold in general, resulting in inferior performance. In order to improve the fully NAR model, multi-pass iterative decoding approaches such as CMLM \cite{Ghazvininejad2019mask-predict} is proposed, which first predicts all of the target words non-autoregressively, and then repeatedly mask out and regenerate the words that the model is least confident about using the masking scheme.

\section{Conclusion}

In this paper, we propose SANAR, a syntax-aware non-autoregressive model to improve the efficiency and accuracy of full-line code completion. We perform an in-depth analysis to explore the dependency among the target tokens, and the results show that the dependency is weaker than the assumption in existing code generation models, suggesting that NAR can fit code generation better. 
Experimental results show that our approach outperforms both NAR and AR baselines, as well as significantly reduces the inference time compared with AR based code completion models. We are the first to apply Non-autoregressive model for generating code. We believe this work represents a significant advance in code generation, which will be beneficial as a building block for many other code generation applications.



\bibliographystyle{named}
\bibliography{ijcai22}

\begin{thebibliography}{}

\bibitem[\protect\citeauthoryear{aiXcoder}{2018}]{aiXcoder}
aiXcoder.
\newblock aixcoder.
\newblock \url{https://www.aixcoder.com/}, 2018.

\bibitem[\protect\citeauthoryear{Allamanis and
  Sutton}{2013}]{Allamanis13Mining}
Miltiadis Allamanis and Charles Sutton.
\newblock Mining source code repositories at massive scale using language
  modeling.
\newblock In {\em {MSR}}, pages 207--216. {IEEE} Computer Society, 2013.

\bibitem[\protect\citeauthoryear{Copilot}{2021}]{Copilot}
Copilot.
\newblock Copilot.
\newblock \url{https://copilot.github.com/}, 2021.

\bibitem[\protect\citeauthoryear{Devlin \bgroup \em et al.\egroup
  }{2019}]{Devlin19bert}
Jacob Devlin, Ming{-}Wei Chang, Kenton Lee, and Kristina Toutanova.
\newblock {BERT:} pre-training of deep bidirectional transformers for language
  understanding.
\newblock In {\em {NAACL-HLT} {(1)}}, pages 4171--4186. Association for
  Computational Linguistics, 2019.

\bibitem[\protect\citeauthoryear{Ghazvininejad \bgroup \em et al.\egroup
  }{2019}]{Ghazvininejad2019mask-predict}
Marjan Ghazvininejad, Omer Levy, Yinhan Liu, and Luke Zettlemoyer.
\newblock Mask-predict: Parallel decoding of conditional masked language
  models.
\newblock In {\em {EMNLP/IJCNLP} {(1)}}, pages 6111--6120. Association for
  Computational Linguistics, 2019.

\bibitem[\protect\citeauthoryear{Gu \bgroup \em et al.\egroup }{2018}]{gu18nat}
Jiatao Gu, James Bradbury, Caiming Xiong, Victor O.~K. Li, and Richard Socher.
\newblock Non-autoregressive neural machine translation.
\newblock In {\em {ICLR} (Poster)}. OpenReview.net, 2018.

\bibitem[\protect\citeauthoryear{Guo \bgroup \em et al.\egroup
  }{2021}]{guo2021learning}
Daya Guo, Alexey Svyatkovskiy, Jian Yin, Nan Duan, Marc Brockschmidt, and
  Miltiadis Allamanis.
\newblock Learning to generate code sketches.
\newblock {\em arXiv preprint arXiv:2106.10158}, 2021.

\bibitem[\protect\citeauthoryear{He \bgroup \em et al.\egroup
  }{2018}]{he2018layer}
Tianyu He, Xu~Tan, Yingce Xia, Di~He, Tao Qin, Zhibo Chen, and Tie-Yan Liu.
\newblock Layer-wise coordination between encoder and decoder for neural
  machine translation.
\newblock In {\em Proceedings of the 32Nd International Conference on Neural
  Information Processing Systems}, pages 7955--7965, 2018.

\bibitem[\protect\citeauthoryear{Hellendoorn and
  Devanbu}{2017}]{hellendoorn2017deep}
Vincent~J Hellendoorn and Premkumar Devanbu.
\newblock Are deep neural networks the best choice for modeling source code?
\newblock In {\em Proceedings of the 2017 11th Joint Meeting on Foundations of
  Software Engineering}, pages 763--773, 2017.

\bibitem[\protect\citeauthoryear{Hindle \bgroup \em et al.\egroup
  }{2016}]{hindle2016naturalness}
Abram Hindle, Earl~T Barr, Mark Gabel, Zhendong Su, and Premkumar Devanbu.
\newblock On the naturalness of software.
\newblock {\em Communications of the ACM}, 59(5):122--131, 2016.

\bibitem[\protect\citeauthoryear{Karampatsis \bgroup \em et al.\egroup
  }{2020}]{karampatsis2020big}
Rafael-Michael Karampatsis, Hlib Babii, Romain Robbes, Charles Sutton, and
  Andrea Janes.
\newblock Big code!= big vocabulary: Open-vocabulary models for source code.
\newblock In {\em 2020 IEEE/ACM 42nd International Conference on Software
  Engineering (ICSE)}, pages 1073--1085. IEEE, 2020.

\bibitem[\protect\citeauthoryear{Kasai \bgroup \em et al.\egroup
  }{2020}]{kasai2020non}
Jungo Kasai, James Cross, Marjan Ghazvininejad, and Jiatao Gu.
\newblock Non-autoregressive machine translation with disentangled context
  transformer.
\newblock In {\em International Conference on Machine Learning}, pages
  5144--5155. PMLR, 2020.

\bibitem[\protect\citeauthoryear{Li \bgroup \em et al.\egroup
  }{2018}]{li2017code}
Jian Li, Yue Wang, Michael~R. Lyu, and Irwin King.
\newblock Code completion with neural attention and pointer networks.
\newblock In {\em {IJCAI}}, pages 4159--4165. ijcai.org, 2018.

\bibitem[\protect\citeauthoryear{Liu \bgroup \em et al.\egroup
  }{2020}]{liu2020self}
Fang Liu, Ge~Li, Bolin Wei, Xin Xia, Zhiyi Fu, and Zhi Jin.
\newblock A self-attentional neural architecture for code completion with
  multi-task learning.
\newblock In {\em Proceedings of the 28th International Conference on Program
  Comprehension}, pages 37--47, 2020.

\bibitem[\protect\citeauthoryear{Ma \bgroup \em et al.\egroup
  }{2019}]{Ma19FlowSeq}
Xuezhe Ma, Chunting Zhou, Xian Li, Graham Neubig, and Eduard~H. Hovy.
\newblock Flowseq: Non-autoregressive conditional sequence generation with
  generative flow.
\newblock In {\em {EMNLP/IJCNLP} {(1)}}, pages 4281--4291. Association for
  Computational Linguistics, 2019.

\bibitem[\protect\citeauthoryear{Qian \bgroup \em et al.\egroup
  }{2021}]{Qian21glat}
Lihua Qian, Hao Zhou, Yu~Bao, Mingxuan Wang, Lin Qiu, Weinan Zhang, Yong Yu,
  and Lei Li.
\newblock Glancing transformer for non-autoregressive neural machine
  translation.
\newblock In {\em {ACL/IJCNLP} {(1)}}, pages 1993--2003. Association for
  Computational Linguistics, 2021.

\bibitem[\protect\citeauthoryear{Raychev \bgroup \em et al.\egroup
  }{2016}]{Raychev16Probabilistic}
Veselin Raychev, Pavol Bielik, and Martin~T. Vechev.
\newblock Probabilistic model for code with decision trees.
\newblock In {\em {OOPSLA}}, pages 731--747. {ACM}, 2016.

\bibitem[\protect\citeauthoryear{Ren \bgroup \em et al.\egroup
  }{2020}]{Ren20Study}
Yi~Ren, Jinglin Liu, Xu~Tan, Zhou Zhao, Sheng Zhao, and Tie{-}Yan Liu.
\newblock A study of non-autoregressive model for sequence generation.
\newblock In {\em {ACL}}, pages 149--159. Association for Computational
  Linguistics, 2020.

\bibitem[\protect\citeauthoryear{Tabnine}{2018}]{tabnine}
Tabnine.
\newblock Tabnine.
\newblock \url{https://www.tabnine.com/}, 2018.

\bibitem[\protect\citeauthoryear{Tu \bgroup \em et al.\egroup
  }{2014}]{tu2014localness}
Zhaopeng Tu, Zhendong Su, and Premkumar Devanbu.
\newblock On the localness of software.
\newblock In {\em Proceedings of the 22nd ACM SIGSOFT International Symposium
  on Foundations of Software Engineering}, pages 269--280, 2014.

\bibitem[\protect\citeauthoryear{Vaswani \bgroup \em et al.\egroup
  }{2017}]{vaswani2017attention}
Ashish Vaswani, Noam Shazeer, Niki Parmar, Jakob Uszkoreit, Llion Jones,
  Aidan~N Gomez, {\L}ukasz Kaiser, and Illia Polosukhin.
\newblock Attention is all you need.
\newblock In {\em Advances in neural information processing systems}, pages
  5998--6008, 2017.

\bibitem[\protect\citeauthoryear{Wang \bgroup \em et al.\egroup
  }{2020}]{wang2020towards}
Wenhan Wang, Sijie Shen, Ge~Li, and Zhi Jin.
\newblock Towards full-line code completion with neural language models.
\newblock {\em arXiv preprint arXiv:2009.08603}, 2020.

\bibitem[\protect\citeauthoryear{Wei \bgroup \em et al.\egroup
  }{2019}]{Wei19Imitation}
Bingzhen Wei, Mingxuan Wang, Hao Zhou, Junyang Lin, and Xu~Sun.
\newblock Imitation learning for non-autoregressive neural machine translation.
\newblock In {\em {ACL} {(1)}}, pages 1304--1312. Association for Computational
  Linguistics, 2019.

\end{thebibliography}

\end{document}